\documentstyle[preprint,floats,aps,psfig]{revtex}
\tightenlines
\begin{document}

\title{Update on Partial-Wave Analysis}

\author{Ron Workman}

\address{Dept. of Physics, The George Washington University, Washington
D.C. 20052\\E-mail: rworkman@gwu.edu}

\maketitle

\begin{abstract}
Partial-wave analysis is one step in a process
connecting experimental measurements to the N$^*$ states we
are studying. Progress has been made in the area of
`model-independent' analysis. However, more model-dependent
approaches are needed to cover broader energy ranges. An
example of the problems faced by these more ambitious analyses
is given.
\end{abstract}

\section{Overview}

The majority of recent partial-wave analyses (PWA),
applied to the study of N$^*$ physics, have focused on 
photo- or electroproduction processes\cite{SAID}. 
In the following, I
will discuss only this topic and further restrict my
comments to pseudoscalar-meson photoproduction.
I will first mention fits that have focused on the 
near-threshold region, where a single resonance is
dominant, and one can argue for a truncated multipole
expansion. Here the goal is a `model-independent' extraction of
amplitudes. Other more ambitious fits have extended over wider
energy ranges where model-dependence is inevitable. As a result,
in such cases, it becomes difficult to 
distinguish between models, with fitted 
parameters, and data-fits (with theoretical constraints). 
The addition of Born contributions will be
discussed as an example illustrating the problems one
encounters in fits to the full resonance region.

Before moving on to examples, it is worth repeating that
the requirements for an `amplitude' analysis and a 
`multipole' analysis are not equivalent. This has been
pointed out in the past\cite{trento}, 
but the distinction is rarely noted in the literature. 
Amplitude analyses have been performed for pion-nucleon and 
nucleon-nucleon scattering. However, in most\cite{kaon} 
reactions of interest, arguments concerning the requirements 
for a `complete experiment' are academic. We will be dealing
with incomplete experimental information and should concentrate
on finding methods which are as model-independent as possible.

In a workshop preceding this meeting, the Baryon Resonance
Analysis Group (BRAG) compared the results of seven different
fits to a benchmark dataset, with an agreed upon method for
handling errors\cite{BRAG}. 
These fits will be refined and repeated over
an expanded dataset (of pion photoproduction data), giving an
objective measure of the model-dependence in multipole 
amplitudes extracted from this reaction.

\section{Near-Threshold Studies}

Much of the recent work on pseudoscalar-meson photoproduction
has focused on the `first bump' region. In pion photoproduction,
this structure is due to the $\Delta (1232)$ resonance. 
Analyses of $\pi^0$ production
(both photo and electroproduction) have used truncated 
(S- and P-wave) multipole expansions with simplifications following
from the dominance of the $M_{1+}^{3/2}$ contribution. In $\eta$ 
photoproduction, one again has a dominant ($E_{0+}^{1/2}$) multipole
near threshold and higher waves are playing a minor role\cite{Tiator}. 
A similar fitting strategy has been applied to 
$\eta '$ photoproduction\cite{Ploetzke},
with less conclusive results.  While the total cross
section displays a near-threshold peak, there is little agreement
on the underlying resonance components\cite{Zhang,Zhao}. 

In $\pi$ and $\eta$ photoproduction, interesting results follow
from fits including measurements 
sensitive to the interference of dominant and
sub-dominant amplitudes. In both cases, the photon asymmetry
($\Sigma$) has played an important role, providing information on
the E2/M1 ratio (for $\pi$) and the 
D$_{13}$(1520) contribution\cite{Tiator}
(for $\eta$). This observable has also had a significant impact on 
pion photoproduction at higher energies\cite{Adamian} and the
`second bump' in $K^+$ photoproduction\cite{Mart}.

These fits have suggested a few new states (in $K^+$ and $\eta'$
photoproduction), and have given new determinations of the
N(1520) and N(1535) photo-decay 
amplitudes (in $\eta$ photoproduction)\cite{Tiator}.
Unfortunately, the new determinations disagree with older ones
obtained from fits to $\pi$ photoproduction data. As a result, the
status of photo-couplings is probably {\em less} 
certain\cite{eta2} than we would have
claimed prior to this renewed experimental effort.

More ambitious analyses, covering broader energy ranges, will be 
required to convincingly establish the existence (or absence) of
the many `missing resonances' predicted by quark models. Clearly,
we must also have {\it mutually consistent} photo-couplings 
from fits to the different reactions. This 
important issue is being
studied by a number of groups\cite{Multi}. 

\section{Fitting the Full Resonance Region}

Most photo-couplings have been determined from fits to $\pi$ 
photoproduction data. This reaction has been extensively studied,
with measurements covering the full resonance region, and at least
an order of magnitude more data than exists for the photoproduction
of other pseudoscalar mesons. In $\pi$ photoproduction
one also has the advantage of
knowing where resonances should occur, given the equally extensive
study of elastic $\pi^{\pm} p$ scattering. 

Moving to other channels, we must rely either on quark model predictions
or much less reliable hadronic information (for example, $\pi N\to \eta N$ 
and $\pi N \to K Y$) for a guide to possible resonance content. The 
total cross section is of little help, generally having a rapid increase
near threshold, and a relatively smooth fall-off at higher energies. As
a result, photo-decay amplitudes will be much more difficult to 
determine (unambiguously) in these reactions. 

One method gaining popularity is the use of a multi-channel formalism
which incorporates information from 
many channels simultaneously. This approach requires
a single formalism capable of describing all included channels. An 
ingredient common to each of these processes is the Born term. In the
following, I will briefly describe how this contribution behaves and
explain why its inclusion is problematic. 

\subsection{The Born Terms: General Features}

The most obvious influence of Born terms in photoproduction is the forward
peak seen in the $\pi^+ n$ differential 
cross section\cite{SAID}. For example, at
a photon energy of 1 GeV, both the data and Born term contribution have
a forward peak and a dip near $t=-\mu^2$. However, away from this forward
peak, the Born contribution has entirely the wrong shape. In addition, the
total cross section from this piece alone exceeds the experimental
value just a few hundred MeV above threshold. This divergence of the Born
contribution from the experimental total cross sections is also seen in
$K^+$ and $\eta '$ photoproduction\cite{Davidson1,Tanabe}. 

At higher energies (10 GeV), a simple Born approximation 
continues to give a qualitative description of the forward peak in
$\pi^+ n$ photoproduction.  This rather unexpected behavior has been 
explained in terms of finite energy sum rules\cite{Dombey}. Thus we have a
Born contribution which is `correct' for the forward peak, but clearly is
problematic for the total cross section. These features, together with
the constraints of gauge invariance and proper analytic structure, can be 
used to assess the model input adopted in fitting data.

\subsection{Taming the Born Contribution}

Born term plus Breit-Wigner fits are adequate to describe
cross section data. However, extrapolation beyond the fitted
energy range tends to
be very unstable. This problem is linked to the behavior of the Born 
component described above.
One simple way to compensate for the Born contribution,
is to multiply by an overall
form factor. This approach clearly violates crossing 
symmetry and reduces the
full angular distribution uniformly, including the forward peak
which requires no suppression. It can even
result in a `resonancelike' Born contribution to
the total cross section\cite{Davidson1}, as demonstrated in a fit to 
$\eta '$ photoproduction.

Some of these problems can be avoided if each strong vertex,
in the Born terms, is modified by a form factor depending on the
off-shell 4-momentum. In this case a
contact term is required to restore gauge invariance.
This can be done in a way which preserves crossing symmetry. 
The effect of these form factor recipes, at the multipole level,
is very different. An overall form factor, $F(s)$, having no angular
dependence, reduces each multipole by the same amount as $s$ increases.
If the recipe of Ref.\cite{Davidson2} is used instead, the effect is
most important for low partial waves, and becomes negligible as the
angular momentum increases. 
However, using 
simple and common (real) form factors, we have found 
that the recipe of Ref.\cite{Davidson2} 
also tends to kill the forward peak seen in $\pi^+ n$.
As a result, it is difficult to see how the use of real form factors
can be justified, based both on formal arguments and in comparison with
general features seen in the data.

\subsection{Results from Dispersion Relations}

All of the above approaches to the Born contribution implicitly separate
the `resonance' and `background' contributions. If one examines this problem
within the context of dispersion relations, an entirely different picture
emerges. In a fixed-t dispersion relation, the real part is given by the
Born term plus a weighted integral over imaginary parts. 
In $\pi$ photoproduction, it has been
shown that the Born term is damped primarily through the inclusion of
$M_{1+}^{3/2}$ in the integral\cite{Engels}. Adding this contribution cancels
the non-forward Born contribution and enhances the forward peak\cite{Barbour}.
Other resonance contributions appear to play a much smaller role, and this
holds far above the energy of the $\Delta(1232)$ resonance.

It is possible to write a Born term with form factors in terms of an
unmodified Born term plus a contact term. 
This contact term must grow with energy to cancel the unmodified Born 
contribution, and therefore represents a significant contribution to 
the real part of the invariant amplitude. 
However, in fits such as the one reported by Crawford at this
meeting, the dispersion integral appears to be well approximated by 
resonance contributions alone (over the resonance region). 

\subsection{SAID versus MAID}

How then are the Born terms handled in the SAID\cite{SAID} and
MAID\cite{Drechsel} fits over the resonance region? In both analyses, the
Born terms are added (without form factors). In MAID, the full Born 
contribution is added to resonance contributions. In SAID, the low
partial waves are parametrized and the high partial waves are 
given by a (real) Born contribution. 

In Figure 13 of Ref.\cite{Drechsel}, a cancellation of the unitarized
Born and $P_{33}$ multipoles was noted in $\pi^+ n$ photoproduction.
We have checked this behavior in SAID and find a very similar 
result. In fact, this simple model is remarkably successful in
reproducing the forward $\pi^+ n$ peak up to the limit of the
MAID analysis (1 GeV). We have also found that this model
results in a qualitative description of very forward (4$^\circ$)
$\pi^\circ p$ differential cross sections over the same energy
region. We did not anticipate this result and haven't found any
specific mention of it in the literature.

In SAID and MAID the Born contribution is damped both by the (K-matrix)
unitarization and a cancellation with the $P_{33}(1232)$ multipoles.
Both contributions are required and important in the observed 
cancellation. 

\section{Conclusions and Further Questions}

Many recent studies have focused on
reactions and kinematic regions with simplifying features.
These have generally been near-threshold fits
where a large signal is present, the final meson is neutral,
and only a few partial-waves are important. 
If our goal is to find the many missing states predicted
by quark models, these exploratory fits  must be
followed by more complete treatments of the full resonance region. 

We have seen that the SAID and MAID fits have features
similar to those mentioned long ago in the context of
fixed-t dispersion relations. Aspects of the forward peaking 
and suppression of Born contributions at non-forward angles
are largely due to the $\Delta (1232)$, with smaller contributions
from other states. These results weigh against approaches
which attempt to separate Born (or generic background) and
resonance contributions, suppressing the Born contribution
through the use of phenomenological form factors. 

Whether the photoproduction of other pseudoscalar mesons
can be treated in an analogous manner is less clear. High
energy $K^+$ photoproduction has a forward dip, rather than
the peak seen in $\pi^+ n$. [Aside: This forward region
was, however, used by Dombey to estimate the $K\Lambda N$ coupling
constant\cite{Dombey}.] Some cancellation of the nucleon
pole term and resonance contributions has been mentioned in
the work of Zhao\cite{Zhao} applied to $\eta '$
photoproduction.

It would be interesting to see how well the various models,
used in fitting data (or partial-wave amplitudes), reproduce
the forward peak seen in $\pi^+ n$ photoproduction. A second
test would involve determining the consistency of each 
approach with the constraints imposed by fixed-t
dispersion relations. Checks of this kind may
help to select the most promising approach to pursue.

\acknowledgments
Thanks are due to R.A. Arndt for modifying SAID specifically
for the NSTAR and BRAG meetings.
Discussions with R.M. Davidson are also gratefully acknowledged. This work
was supported in part by the U.S. Department of Energy Grant
DE-FG02-99ER41110 and a contract from Jefferson Lab, which is 
operated by the Southeastern Universities Research Association
under the U.S. Department of Energy Contract DE-AC05-84ER40150.

\eject


\begin{thebibliography}{99}
\bibitem{kaon}Kaon photoproduction experiments proposed
at JLab may provide a `nearly complete' set of observables.
\end{thebibliography}
\end{document}